\documentclass[aps,prm,preprint,superscriptaddress,showpacs,floats]{revtex4-2}
\usepackage{graphicx}
\usepackage{amsfonts}
\usepackage{amsmath}
\usepackage{amssymb}
\usepackage{exscale}
\usepackage{afterpage}				
\usepackage{lineno}
\usepackage{xcolor}
\usepackage{natbib}

\begin{document}

\author{Olivier Copie}
\email{olivier.copie@univ-lorraine.fr}
\affiliation{Universit\'e de Lorraine, CNRS, IJL, F-54000 Nancy, France}
\author{Julien Varignon}
\affiliation{CRISMAT, CNRS, Normandie Universit\'e, ENSICAEN, UNICAEN, F-14000 Caen, France}
\author{Ingrid Cañero Infante}
\affiliation{Institut des Nanotechnologies de Lyon, CNRS ECL INSA UCBL CPE, F-69621 Villeurbanne, France}
\author{Mariam Martirosyan}
\affiliation{Universit\'e de Lorraine, CNRS, IJL, F-54000 Nancy, France}
\author{Fadi Choueikani}
\affiliation{Synchrotron SOLEIL, L'Orme des Merisiers, F-91190 Saint-Aubin, France}
\author{Philippe Ohresser}
\affiliation{Synchrotron SOLEIL, L'Orme des Merisiers, F-91190 Saint-Aubin, France}
\author{Pierre-Eymeric Janolin}
\affiliation{Université Paris-Saclay, CentraleSupélec, CNRS, Laboratoire SPMS, 91190, Gif-sur-Yvette, France}
\author{Alain Pautrat}
\affiliation{CRISMAT, CNRS, Normandie Universit\'e, ENSICAEN, UNICAEN, F-14000 Caen, France}
\author{Adrian David}
\affiliation{CRISMAT, CNRS, Normandie Universit\'e, ENSICAEN, UNICAEN, F-14000 Caen, France}
\author{Philippe Ghosez}
\affiliation{Theoretical Materials Physics, Q-MAT, Université de Liège, B-4000 Liège, Belgium}
\author{Wilfrid Prellier}
\affiliation{CRISMAT, CNRS, Normandie Universit\'e, ENSICAEN, UNICAEN, F-14000 Caen, France}
\title{Tailoring non-collinear magnetism and $3d-4f$ exchange interactions in RVO$_3$ epitaxial thin films}
%
%
%
%\date{\today}
%
%
\begin{abstract}
In orthorhombic perovskite oxides (RMO$_3$), substituting R$^{3+}$ rare-earth cations tailors the spin, orbital, and charge degrees of freedom of the central M$^{3+}$ transition metal cations through lattice distortions. In turn, these modify also the surrounding environment of R$^{3+}$. When both R$^{3+}$ and M$^{3+}$ exhibit magnetic properties, phenomena such as spin reorientation and magnetization reversal can occur. In fact, the underlying exchange interactions between M-$3d$ spins and R-$4f$ magnetic moments enrich the multifunctional character of RMO$_3$, particularly when combined with structural distortions. They play a crucial role in achieving appealing properties such as robust magnetoelectricity with non-collinear magnetic orders. Here, we explore the exchange coupling in epitaxial PrVO$_3$ thin films, selectively probing the magnetism of cation sublattices, and uncovering simultaneous V$^{3+}$ $3d$ spin reorientation and Pr$^{3+}$ $4f$ magnetization reversal using spectroscopy techniques. By strain engineering, we manipulate the lattice distortions to rationalize their role in coupling $3d$ spins and $4f$ magnetic moments. Theorectical calculations show that octahedral rotations and Jahn-Teller distortions act as tuning mechanisms, promoting competition between orbital and spin orders. The observed coupling between magnetic cations and lattice distortions can be extended to other  orthorhombic RMO$_3$ systems, advancing the understanding of controlling spins in engineered perovskite heterostructures and superlattices.
\end{abstract}
\maketitle
\newpage
%
%\linenumbers
%
\section{Introduction}
Understanding the complexity of transition metal oxides (TMO) requires the consideration of the many degrees of freedom exhibited by the electrons in the solid. Additionally, the lattice structure can influence significantly the physical properties, and even a slight change in the crystal structure results in notable changes in the functional properties~\cite{Goodenough2004}. Over decades, extensive research has been dedicated to comprehending the intricate interplay among spin, orbital and charge orders in the perovskite oxide compounds~\cite{Imada1998,Tokura2000}. These concurrently active degrees of freedom give rise to a multitude of correlated electronic states that underlie the richness of the functionalities~\cite{Zubko2011}. This diversity fuels a strong interest in delving into the fundamentals and the potential applications of perovskite-related materials~\cite{Coll2019}. 
Due to the perovskite structure's inherent ability to distort, RMO$_3$ can accommodate many atoms of the periodic table, with different atomic radii. When both R$^{3+}$ and M$^{3+}$ ions are magnetic, intriguing magnetic behavior are reported~\cite{Bousquet2016}, such as the change of the spin direction along different crystallographic axes or the inversion of the net magnetization of the crystal. These phenomena, known as spin reorientation and magnetization reversal, hold significant importance for applications~\cite{Fert2024}. Additionally, complex non-collinear magnetic orders may manifest, breaking inversion symmetries and inducing an electric polarization~\cite{Bousquet2016, Tokura2010}, resulting in a robust magnetoelectric response~\cite{Kimura2003}. This interplay of magnetic and electric properties further intensifies the interest in these materials~\cite{Manipatruni2019}. 

Here, we study the magnetic coupling in epitaxial RVO$_3$ thin films between R$^{3+}$ (R = Pr) and V$^{3+}$. We selectively probed the magnetic behavior of the cation sublattices and observed the simultaneous V$^{3+}$ $3d$ spin reorientation and Pr$^{3+}$ $4f$ magnetic moments reversal, both dependent on the temperature and magnetic field. Through strain engineering, we precisely tailored the lattice structure, which in turn modifies the lattice distortion such as the oxygen octahedra rotation, involved in the couplings between $3d$ and $4f$ magnetic moments~\cite{Zhao2016b}. This exchange strictive interaction, beside the understanding of the robust magnetoelectricity in orthorhombic perovskite~\cite{Zhao2016a}, is important for designing artificial multiferroic superlattices~\cite{Rondinelli2012,Varignon2015, Spaldin2007}.   

To accommodate the mismatch between R$-$O and M$-$O bond lengths, RMO$_3$ commonly adopts a $Pbnm$ orthorhombic structure, characterized by cooperative oxygen octahedron rotations, as illutrated in Figure~\ref{Fig1}(a) featuring the tilted MO$_6$ octahedra (in blue) with the central M$^{3+}$ ion. The rotations also modify the local environment around the R$^{3+}$ cation, which is surrounded by 12 oxygen atoms forming a cuboctahedron, as depicted in Figure~\ref{Fig1}(a) (in green). Additionally, due to the change of the R$-$O covalency, the rotations of the MO$_6$ octahedra lead to a shift of R$^{3+}$ ions from their position in cubic symmetry~\cite{Goodenough1971}, being a key parameter of hybrid improper ferroelectricity~\cite{Bousquet2008,Benedek2011}. All these atomic displacements resulting from structural changes alter the orbital occupancy and electron transfer between neighboring atoms, potentially giving rise to complex orbital and spin ordered phases in RVO$_3$ compounds and others~\cite{Miyasaka2003,Dagotto2005,Sage2007, Radhakrishnan2024}. 

While the $3d$ delocalized electrons are sensitive to the strength of the crystal field, the $4f$ localized electrons undergo a large spin-orbit coupling (SOC) that enhances the magnetic anisotropy. Therefore, beyond merely considering the M$^{3+}$ ion and the structural distortion of its closest neighborhood, the role of the magnetic exchange between the M$^{3+}$ and R$^{3+}$ sublattices should be investigated~\cite{Zhao2016b,Sasani2021}. We point out that, recently, the effects of rare earth magnetism and $4f$ electrons have also been the focus of attention with regard to the superconducting properties of the infinite-layer RNiO$_2$ system~\cite{Li2019, Wang2023}.

RVO$_3$ compounds have emerged as an intriguing material, drawing significant interest to better understand the properties of TMO. Noteworthy features include a complex yet prototypical interplay between spin, orbital, and charge orders, coupled with crystalline lattice distortions, such as relatively weak Jahn-Teller (JT) distortions~\cite{Miyasaka2003}. In contrast to Mn$^{3+}$ $e_g$ electrons in RMnO$_3$, the JT distortions of the two $t_{2g}$ electrons at V$^{3+}$ sites subtly bias the orbital ordering~\cite{Zhou2008}. The cooperative VO$_6$ rotations are described by an $a^-a^-c^+$ tilt pattern according to Glazer's notation~\cite{Glazer1972} (out-of-phase rotation $\varphi_{x}^{-}$ ($\varphi_{y}^{-}$) along the $[100]_{\rm c}$ ($[010]_{\rm c}$) axis , and in-phase rotation $\varphi_{z}^{+}$ along the $[001]_{\rm c}$ axis, c for cubic)~\cite{Sage2007}. This JT distortion, consisting in long and short V$-$O bonds length along the $[100]_{\rm c}$ and $[010]_{\rm c}$ pseudo-cubic directions~\cite{Zhou2005}, allows lifting the degeneracy between the V $t_{2g}$ levels. The two electrons per V$^{3+}$ sites are then ideally located either in the $d_{xz}$ or in the $d_{yz}$ orbitals, $d_{xy}$ being always occupied. As the temperature decreases, RVO$_3$ undergo a structural phase transition from orthorhombic $Pbnm$ to monoclinic $P2_1/b$~\cite{Miyasaka2003,Sage2007}. Concurrently, the cooperative JT distortions promote a G-type orbital order (G$_{\rm{OO}}$) of V $t_{2g}$ orbitals, as presented in Figure~\ref{Fig1}(b), at the temperature T$_{\rm{OO}}$. At lower temperature, superexchange interactions then stabilize a C-type antiferromagnetic spin order (C$_{\rm{SO}}$) of the V$^{3+}$ spins at the temperature T$_{\rm{SO}}$, as shown in Figure~\ref{Fig1}(b), also referred as N\'eel temperature. As R$^{3+}$ ionic size decreases, the VO$_6$ rotations increase and T$_{\rm{SO}}$ decreases monotonously with the V $3d~-$~O $2p$ orbitals overlap reduction, as for RCrO$_3$ or RFeO$_3$~\cite{Zhou2010}. For small R$^{3+}$ ionic radii, a structural transition back to $Pbnm$ occurs, accompanied by a spin/orbital reorderings from C$_{\rm{SO}}$/G$_{\rm{OO}}$ to G$_{\rm{SO}}$/C$_{\rm{OO}}$~\cite{Miyasaka2003}.    

\section{Experimental and theoretical Methods}
40~nm-thick PrVO$_3$ thin films were deposited onto (001)$_{\rm c}$-oriented SrTiO$_3$ substrates by PLD, firing on a polycrystalline PrVO$_4$ target. The latter was prepared via standard solid-state reaction from a stoichiometric Pr$_6$O$_{11}$ and V$_2$O$_5$ powders, subsequently pressed into pellets and sintered in air at 1100~$^{\circ}$C for 48 hours. The KrF excimer laser ($\lambda=248$~nm) was focused on the target with a repetition rate and a fluence set at 2~Hz and 2~J.cm$^{-2}$, respectively. Prior to deposition, the substrates underwent cleaning in acetone and propanol. The deposition was carried out within a growth temperature range of 500$^{\circ}$C to 800$^{\circ}$C, maintaining the growth pressure fixed to 8$\times$10$^{-6}$~mbar. The lattice parameters were determined using X-ray diffraction with a Malvern PANalitycal X’pert Pro MRD diffractometer, with a monochromatic Cu $K\alpha_1$ radiation ($\lambda = 1.54056$~\AA).

The macroscopic magnetization was measured in DC mode using a superconducting quantum interference device (SQUID) magnetometer. A magnetic field \textbf{B} was applied parallel and perpendicular to the film surface, employing a Quantum Design MPMS ($\pm$50~kOe) and MPMS3 ($\pm$70~kOe), respectively. The measured magnetization was corrected by subtracting the substrate diamagnetic contribution.
The FC (Field Cooled) --ZFC (Zero Field Cooled) protocol consisted in cooling down the sample with $|\textbf{B}|=5$~kOe, then measuring the magnetization as function of the temperature up to 300~K with $|\textbf{B}|=50$~Oe, and cooling down the sample without applied magnetic field, then measuring the magnetization as function of the temperature up to 300~K with $|\textbf{B}|=50$~Oe. 

Soft X-ray absorption spectroscopy (XAS) and X-ray magnetic circular dichroism (XMCD) were performed at the DEIMOS beamline of the French Synchrotron Facility SOLEIL~\cite{Ohresser2014}. XAS is a powerful technique due to its chemical/orbital selectivity and sensitivity to the crystalline environment of the absorber atom. XMCD provides information such as the spin and orbital magnetic moments without the substrate's contribution. Consequently, we selectively probed the roles of the V$^{3+}$ and the Pr$^{3+}$ magnetic sublattices. 
The V-L$_{2,3}$ edges correspond to the excitation of photoelectrons from the spin-orbit split $2p_{1/2}$ and $2p_{3/2}$ levels to spin-split $3d$ level and the Pr-M$_{4,5}$ edges correspond to the $3d_{3/2}$ and $3p_{5/2}$ to $4f$ transitions. Additionally, spectra at the O-K edge were recorded, corresponding to the excitation of photoelectrons from the O $1s$ orbital to O $2p$ orbitals, which are hybridized with the empty orbital states of the V and Pr cations. 
The measurements were carried out using the total electron yield (TEY) mode. The temperature ranged from 5~K to 300~K, with the applied magnetic field of $\pm$65~kOe directed along the X-ray beam. 
Eight absorption spectra were acquired across the V-L$_{2,3}$ and O-K (from 510 to 550~eV) and Pr-M$_{4,5}$ (from 920 to 970~eV) edges, alternating opposite magnetic field directions and left/right circular (CL/CR) ligth polarization, and then averaged to obtain the $\mu_{\rm{CR}}$ and $\mu_{\rm{CL}}$ spectra. Subsequently, the XMCD spectra were calculated as the difference  $0.5\times(\mu_{\rm{CR}}-\mu_{\rm{CL}})$. The probing depth is approximately 6~nm and 15~nm around 550~eV and 950~eV, respectively. 
The first XAS and XMCD spectra from 0 to 65~kOe were measured after zero field cooling. The magnetic field-dependent XMCD loops at the V-L$_3$ and Pr-M$_4$ edges peaks were obtained by measuring, at each field, the difference between the intensity for CR and CL light polarization, normalized by the intensity before the absorption edge. 
The energy resolution of the XAS and XMCD experiments were around 0.06~eV and 0.15~eV across the V-L$_{2,3}$/O-K and the Pr-M$_{4,5}$, respectively. From the XMCD spectra, quantitative information such as the orbital ($\langle\rm L_{\rm{z}}\rangle$) and the spin (2$\langle\rm S_{\rm{z}}\rangle$) moments can be extracted using the so-called sum-rules~\cite{Thole1992,Carra1993}. Unfortunately, the large overlap between the V-L$_2$ and L$_3$ edges complicates the application of the rules for the determination of $\langle\rm S_{\rm{z}}\rangle$. As detailed later in the text, the spin sum rule is also invalid for light rare earths such as Pr. Therefore, we have only applied the orbital sum rule Pr-M$_{4,5}$ edges. Hence, $\langle\rm L_{\rm{z}}\rangle$~$=(4l+2-n)\times q/r$ with $q=\int_{\rm{M}_4+\rm{M}_5}(\mu_{\rm{CR}}-\mu_{\rm{CL}})dE$ and $r=\int_{\rm{M}_4+\rm{M}_5}0.5\times (\mu_{\rm{CR}}+\mu_{\rm{CL}})dE$. The transitions into the higher unoccupied states or into the continuum have been corrected in the XAS spectra by a two step-like function. In the case of Pr$^{3+}$ ($4f^2$), the angular moment $l$ is 3, and the occupation number $n$ is 2, \textit{i.e.} the number of holes is 12.

The density functional theory (DFT) calculations were carried out using the VASP package \cite{Kresse1993,Kresse1996}. A 4$\times$4$\times$4 $k$-point grid was utilized for the 40 atoms $Pbnm$ ($P2_1/b$) phases and their respective symmetry lowered $P2_1/m$ ($P{\bar 1}$) film structures. The energy cutoff was set to 500 eV. The convergence of the self-consistent field was achieved until the energy difference between consecutive steps was lower than 10$^{-9}$ eV. The geometry relaxations were conducted until forces were lower than 0.1~meV/\AA. We used the Projector Augmented-Wave method (PAW) with the following valence electrons: $5s^25p^66s^0$ for Pr$^{3+}$, the two extra $4f^2$ electrons are either frozen in the pseudopotential or treated explicitly ; $3s^23p^64s^03d^2$ for V$^{3+}$) ; $2s^22p^6$ for O$^{2-}$. By default, the Pr $4f$ electrons are frozen in the pseudopotential, while their explicit treatment is only considered when addressing the magnetism at Pr site. We employed the PBE functional revised for solids (PBEsol)~\cite{Perdew2008}. To account for the strong correlation of the V$^{3+}-t_{2g}$ electrons \cite{Liechtenstein1995}, the effective U parameter acting on V $d$ orbitals is fixed to U$\rm_{eff}=3.5$~eV, a value previously shown to reproduce the RVO$_3$ phase diagram~\cite{Varignon2015} as well as the experimental trend of the Néel temperature of PrVO$_3$ experiencing epitaxial and/or chemical strain~\cite{Copie2017, Kumar2019a}. Unless stated, DFT calculations are performed without Pr $4f$ electrons. For calculations involving Pr $4f$ electrons, delocalization errors inherent to the implementation of DFT being larger for $4f$ orbitals than for $3d$ orbitals, we opted for an effective parameter U$_{4f}$ of 5 eV acting on Pr $4f$ states. This value appears as a fair choice reproducing a more stable V-C$_x$C$_z$ Pr-C$_x$C$_z$ magnetic solution than a V-G$_y$ Pr-C$_x$C$_z$ in the bulk PrVO$_3$ [$\rm\Delta E=-5.65$~meV (f.u)$^{-1}$], in agreement with experimental observations (see Supplemental Material~\cite{SM}). The spins were treated at the collinear and non-collinear levels, and two different antiferromagnetic spin orderings were used, namely C$_{\rm{SO}}$ and G$_{\rm{SO}}$ spin orders for V. The insulating C$_{\rm{SO}}$/G$_{\rm{OO}}$ ground state of the bulk PrVO$_3$ is correctly predicted by preliminary DFT calculations. In the case of PrVO$_3$ on (001)$_{\rm c}$-oriented SrTiO$_3$, the $[001]$ and $[1\bar{1}0]$ orthorhombic axes were aligned along the in-plane cubic axes of substrate while the $[110]$ axis was forced to be orthogonal. PrVO$_3$ then adopts a $P2_1/m$ symmetry with a C$_{\rm{OO}}$/G$_{\rm{SO}}$ ground state. The strain was treated as an uniaxial constraint on the $[110]$ direction of the film and only atomic positions were relaxed. 

\section{Results and discussion}
We have deposited PrVO$_3$ epitaxial films by pulsed laser deposition (PLD) on (001)$_{\rm c}$-oriented SrTiO$_3$ cubic substrate (see above). The orthorhombic $Pbnm$ symmetry is lowered to the monoclinic $P2_1/m$ one, the long $c$-axis being clamped by one of the in-plane substrate cubic directions, as the in-phase octahedral rotation of PrVO$_3$~\cite{Copie2013,Kumar2019a}. The uniaxial strain along the $[110]$ direction was then tailored by controlling the oxygen exchange between the substrate and the film during the growth~\cite{Copie2017}, increasing the unit cell volume V$_{\rm{UC}}$. We point out that the (110)-oriented films maintain some structural degrees of freedom that enable them to undergo the symmetry change compatible with the spin orderings. The structural characterizations of the heterostructures, including measurements at low temperatures, are given in the Supplemental Material~\cite{SM}.

\subsection{Spin reorientation.} In Figure~\ref{Fig1}(c)--(f), we depict the magnetization as a function of an applied in-plane magnetic field \textbf{B} after zero field cooling. At 10~K, a hysteretic behavior with a coercive field (H$_{\rm{C}}$) of approximately 30~kOe is observed. At remanence, the net magnetization should result from the canting of the V$^{3+}$ spins due to the antisymmetric exchange interaction $\textbf{D}_{ij}.(\textbf{S}_i\times\textbf{S}_j)$, where $\textbf{D}_{ij}$ is the Dzyaloshinskii-Moriya term and $\textbf{S}_{i(j)}$ the spin moment at site $i(j)$. In Figure~\ref{Fig1}(c), the net magnetization linearly increases with \textbf{B} from 10~kOe up to the saturation, typical of \textbf{B} perpendicular to the easy axis of the antiferromagnetically coupled sublattices (small V$_{\rm{UC}}$). Below 10~kOe, a zero net magnetization plateau is observed, widening with increasing V$_{\rm{UC}}$. In Figure~\ref{Fig1}(f), the plateau extends to 30~kOe. Above, the magnetization rises rapidly until saturation, representing \textbf{B} being parallel to the easy axis before a metamagnetic transition clearly seen for larger V$_{\rm{UC}}$. The magnetic hysteresis evolution [Figure~\ref{Fig1}(c)--(f)] suggests a competition or coexistence of the two spin orientations with varying V$_{\rm{UC}}$. In the bottom panel of Figure~\ref{Fig1}(k), the saturation magnetization (M$_{\rm{S}}$) and remanence (M$_{\rm{R}}$) increases as V$_{\rm{UC}}$ decreases. This implies an energetically more favorable orientation of \textbf{B} relative to the spins' easy axis, specifically perpendicular in the case of antiferromagnets with high anisotropy energy compared to the exchange energy of the magnetic sublattices. Consequently, modifying the PrVO$_3$ unit cell strain provides a control of the anisotropy energy and may influence different spin orientations. 

\subsection{Magnetization reversal.} In Figure~\ref{Fig1}(g)--(j), we present the temperature-dependence of the magnetization following the FC -- ZFC protocol (see above). For the FC curve, the magnetic signal increases as the temperature decreases, corresponding to the ordering of the V$^{3+}$ spins at T$_{\rm{SO}}$. For the ZFC, it turns negative below T$_{\rm{SO}}$ down to 10~K as seen in the insets. In details, the magnetization initially increases at T$_{\rm{SO}}$, then it reverses, crosses zero few degrees below and becomes more negative as V$_{\rm{UC}}$ increases. This behavior is archetypal of the existence of two distinct magnetic sublattices with equal magnetic moments compensating each other at the so-called compensation temperature (T$_{\rm{COMP}}$) until the magnetic moment of a sublattice prevails at lower temperature. In Figure~\ref{Fig1}(k), we display T$_{\rm{SO}}$ and T$_{\rm{COMP}}$ as a function V$_{\rm{UC}}$. The non-monotonous evolution of T$_{\rm{SO}}$ has been elucidated by considering the opposing behavior of intra-plane and inter-plane magnetic superexchange interactions between nearest neighbors V$^{3+}$ sites~\cite{Copie2017}. Although T$_{\rm{COMP}}$ behaves similarly, the difference with T$_{\rm{SO}}$ increases monotonoulsy with V$_{\rm{UC}}$, suggesting a structural bias between magnetic sublattices at different crystallographic sites. In RVO$_3$, the temperature-induced magnetization reversal has been attributed to simultaneous switching of the spin and orbital orders in YVO$_3$~\cite{Ren1998} or spin imbalance between two inequivalent V$^{3+}$ sublattices in SmVO$_3$ and NdVO$_3$~\cite{Kimishima2000,Kimishima2005}. Furthermore, the magnetic moment carried by R$^{3+}$ can influence the magnetic phase diagram of RVO$_3$ as observed in bulk by neutron diffraction~\cite{Reehuis2006,Reehuis2011,Reehuis2016}. We see a distinct jump at zero magnetic field in the magnetic hysteresis loops depicted in Figure~\ref{Fig1}. This is attributed to the paramagnetic contribution of isolated Pr$^{3+}$ cations owing to the overoxidation of the film surface after air exposure and the presence of V$^{4+}$ cations~\cite{Kumar2019b}. This effect is commonly observed at the surface of TMO~\cite{Cao2015,Aeschlimann2018}. But intuitively, this points the existence of a magnetic coupling between Pr$^{3+}$ and V$^{3+}$ in stoichiometric PrVO$_3$. 

\subsection{Ferrimagnetic coupling.} To investigate the magnetic and structural couplings in the PrVO$_3$ thin films, we have selectively probed the magnetic properties of the V$^{3+}$ and Pr$^{3+}$ sublattices by using XAS and XMCD (if not specified, the measurements are performed on the film $\#4$). The experimental geometry is illustrated in Figure~\ref{Fig2}(a) and shows the parallel alignment of the incoming polarized light, the magnetic field \textbf{B} and the $[110]$ direction, being the quantization axis. In Figure~\ref{Fig2}(b) and (c), the XAS spectra measured at 5~K, with $|\textbf{B}|=65$~kOe, are presented for the V-L$_{2,3}$ and Pr-M$_{4,5}$ edges, respectively. In comparison with spectra of the literature, the V-L$_{2,3}$ spectral line shape reveals a mixture of V$^{3+}$ (around 516~eV) and V$^{4+}$ around (around 517.5~eV)~\cite{Wadati2009} at the very surface (see Figure S3 and Figure S4 in ref.~\cite{SM}), while the Pr-M$_{4,5}$ edges fine structures are characteristic of Pr$^{3+}$ ions only and in very good agreement with atomic calculations considering a quantum number $j=4$ ground state~\cite{Goedkoop1988}. 
Figure~\ref{Fig2}(b) and (c) display also the corresponding XMCD spectra. A small dichroic signal at the V-L$_{2,3}$ edges indicates a net magnetic moment, along the $[110]$ direction, due to the canted V spins. We observe also a large dichroic signal at the Pr-M$_{4,5}$. According to our experimental geometry the Pr and V total magnetic moments, projected along the $[110]$ direction, are parallel.
Figure~\ref{Fig2}(d) and (e) depict the magnetic field dependence of the XMCD signal at the V-L$^{(1)}_{3}$ edge (516~eV) and the M$_{4}$ (954~eV) edges. The similar shape and width of the hysteresis loops indicate the mutual follow-up and rotation of the Pr and V moments during the field switching. Within experimental accuracy, the deduced coercive fields are fairly similar: 8.7$\pm$1~kOe (V-L$_{3}$) and 6.3$\pm$0.5~kOe (Pr-M$_{4}$), providing a direct evidence of the existence of two coupled magnetic sublattices in PrVO$_3$. We point out that the hysteretic behavior is only observed when measured at the V-L$^{(1)}_{3}$ edge, whereas a reversible slope is observed at the V-L$^{(2)}_{3}$ edge and confirms the isolated and paramagnetic character of the V$^{4+}$ cations at the very surface in Figure~\ref{Fig2}(d).
In Figure~\ref{Fig2}(d) and (e), the linear and reversible XMCD loops at the V-L$^{(1)}_{3}$ and Pr-M$_{4}$ edges at 125~K demonstrate the paramagnetic behavior above T$_{\rm{SO}}$, further supporting the intrinsic magnetic coupling of the V$^{3+}$ and Pr$^{3+}$ sublattices in PrVO$_3$. 
Although, the so-called sum rules are powerful for the quantitative analysis of XMCD spectra~\cite{Thole1992,Carra1993}, care must be taken when using them on early transition metals, especially for the spin momentum $\langle\rm S_{\rm{z}}\rangle$. For the orbital momentum $\langle\rm L_{\rm{z}}\rangle$, as it is proportional to the total integrated spectrum, the zero integrated V-L$_{2,3}$ dichroic signal indicates the orbital moment is zero (see Figure S5 in ref.~\cite{SM}). Despite, we consider the shaded area under V-L$^{(2)}_{1}$ in Figure~\ref{Fig2}(b) as representative of the V$^{3+}$ total moment. The area is denoted as $\cal{A}_{\rm{V^{3+}}}$. For Pr$^{3+}$, only the orbital sum rule can again be applied. Indeed, strong Coulomb interactions between the $4f$ electrons and the $3d$ core hole lead to a significant deviation of the $\langle\rm S_{\rm{z}}\rangle$ calculated values~\cite{Schille1994,Teramura1996}, which prohibits considering it here. Nevertheless, as the Pr$^{3+}$ shape is well reproduced by atomic calculations in a $j=4$ ground state~\cite{Goedkoop1988}, we assume the orbital moment dominates and $\langle\rm L_{\rm{z}}\rangle$ serves, from here, as a proxy catching the magnetic behavior of the $4f^2$ electrons in PrVO$_3$. Using the orbital sum rule, we determined $\langle\rm L_{\rm{z}}\rangle=-1.11~\mu_B$ at.$^{-1}$ for Pr$^{3+}$ (see above), at 5~K, $|\textbf{B}|=65$~kOe, and maximum when \textbf{B} is along $[110]$ (see Figure S4 in ref.~\cite{SM}). As Pr$^{3+}$ cation has less than half-filled shells, $\langle\rm L_{\rm{z}}\rangle$ and $\langle\rm S_{\rm{z}}\rangle$ combine as 2$\langle\rm S_{\rm{z}}\rangle$ reduces the dominating contribution of $\langle\rm L_{\rm{z}}\rangle$ to the total magnetic moment. Consequently, the total magnetization of PrVO$_3$ is the result of a balance between V$^{3+}$ spins and Pr$^{3+}$ orbital and spin moments.

\subsection{4f polarized moments and 3d ordered spins.} Figure~\ref{Fig3}(a) and (b) illustrate the XMCD spectra at the V-L$_{\rm{2,3}}$ and Pr-M$_{\rm{4,5}}$ edges, respectively, with $|\textbf{B}|=65$~kOe, across varying the temperature. The V$^{3+}$ dichroic signal exhibits a rapid initial drop, followed by a smooth decrease upon heating to 300~K. The Pr$^{3+}$ dichroic signal gradually decreases up to 300~K. In Figure~\ref{Fig3}(c), we show the integrated dichroic signal $\cal{A}_{\rm{V^{3+}}}$ and $\langle\rm L_{\rm{z}}\rangle$ against the temperature. Below 100~K, $\cal{A}_{\rm{V^{3+}}}$ shows a notable increase, and a comparable trend is observed for $\langle\rm L_{\rm{z}}\rangle$ amplitude. This agrees with the onset of the V$^{3+}$ spin ordering at T$_{\rm{SO}}=81$~K [see Figure~\ref{Fig1}(j)]. The inset in Figure~\ref{Fig3}(c) displays the inverse of $\cal{A}_{\rm{V^{3+}}}$ as a function of the temperature. The inverse of $\cal{A}_{\rm{V^{3+}}}$ versus temperature conforms to a Curie-Weiss law $\propto(T+\theta)/C_{\rm{V^{3+}}}$, where $C_{\rm{V^{3+}}}$ is the Curie constant of the transition metal sublattice, and $\theta$ is the Weiss temperature being negative and characterizing the resistance to parallel alignment of the V$^{3+}$ spins. The Pr$^{3+}$ contribution to the total magnetization is found to rise around T$_{\rm{SO}}$. Moreover, we observe that the dichroism at the Pr-M$_{\rm{4,5}}$ edges persists, as shown in the inset of Figure~\ref{Fig3}(b), without applied magnetic field up to 40~K. Both temperature-dependence and persistence of the Pr-M$_{\rm{4,5}}$ edges dichroism further agree with an exchange interaction between the ordered V$^{3+}$ spins and the paramagnetic Pr$^{3+}$ moments, yielding a polarized Pr$^{3+}$ sublattice.
The calculated momenta yield of $\langle\rm L_{\rm{z}}\rangle$ is $-0.12~\mu_B$ at.$^{-1}$ and $0.13~\mu_B$ at.$^{-1}$ at 5~K and 40~K, respectively. Eventually, this reversal shows that the balance between the magnetic sublattices is temperature dependent and may be canceled out at T$_{\rm{COMP}}$ as observed in Figure~\ref{Fig1}.

\subsection{Non-collinear magnetism.} In Figure~\ref{Fig4}(a) and (b), we present the XMCD spectra recorded at 5~K following ZFC for the V-L$_{\rm{2,3}}$ and Pr-M$_{\rm{4,5}}$ edges, respectively. The dichroic V$^{3+}$ and Pr$^{3+}$ signals exhibit an increase with changing \textbf{B}. The insets compare the XMCD spectra measured at zero field, after ZFC (light brown) and after saturation (dark brown). While the V-L$_{2,3}$ XMCD signal  at remanence indicates a net moment carried by V$^{3+}$, the one after ZFC corresponds to the derivative of the XAS spectra (not shown), indicating there is no sizable net moment projected along the $[110]$ direction after ZFC. The Pr-M$_{4,5}$ signal is dichroic both after ZFC and at remanence, but with opposite signs at the Pr-M$_{4}$ edge and smaller intensity after ZFC. Since the maximum dichroic intensity occurs when the magnetization direction (quantization axis) is aligned with the photon spin direction~\cite{Stohr1999}, the smaller intensity after ZFC indicates a resulting magnetic moment either very small or moving away from the $[110]$ direction. In Figure~\ref{Fig4}(c), we present the \textbf{B} dependence of $\cal{A}_{\rm{V^{3+}}}$. It displays a metamagnetic-like transition with increasing \textbf{B}. $\cal{A}_{\rm{V^{3+}}}$ increases gradually between 0 and $\sim$20~kOe, stabilizes between 20 and 40~kOe, and then increases proportionally to \textbf{B}. We propose the following scenario, considering the projection of the magnetic moment along the direction of \textbf{B}. (1) after ZFC, the V$^{3+}$ spins are antiparallel, resulting in no sizable magnetization, as depicted in Figure~\ref{Fig4}(c). As \textbf{B} is perpendicular to the spin easy axis (red dashed line), the spins of the V$^{3+}$ sublattice cant in the direction of \textbf{B}, gradually building up a net magnetization between (1) and (2). The plateau between (2) and (3) suggests that the anisotropy energy is not compensated by \textbf{B}, which occur between (3) to (4) where the V$^{3+}$ spins cant further as \textbf{B} increases. This implies the rotation of the spin easy axis, becoming parallel to \textbf{B}. As shown in Figure~\ref{Fig1}(b), the spin axes of G$_{\rm{SO}}$ and C$_{\rm{SO}}$ are perpendicular. We note that first-principles calculations using density functional theory (DFT) have shown that at the collinear level (see above), PrVO$_3$ adopts a $P2_1/m$ symmetry with a G$_{\rm{SO}}$ ground state for the V$^{3+}$ spins without \textbf{B}. Therefore, we rationalize the reorientation of the V$^{3+}$ spins with \textbf{B} as a spin ordering transition from G$_{\rm{SO}}$ to C$_{\rm{SO}}$. To tackle the magnetic behavior of the Pr$^{3+}$ sublattice in the presence of a magnetic field, we present the field-dependence of $\langle\rm L_{\rm{z}}\rangle$ in Figure~\ref{Fig4}(c). At $|\textbf{B}|=0$, $\langle\rm L_{\rm{z}}\rangle=0.18\mu_{B}$ at.$^{-1}$ and at 10~kOe, it reverses to $-0.490\mu_{B}$ at.$^{-1}$, remaining globally proportional to \textbf{B}. At remanence $\langle\rm L_{\rm{z}}\rangle$ remains negative ($-0.12~\mu_B$ at.$^{-1}$). It seems that the V$^{3+}$ reordering affects the Pr$^{3+}$ magnetic moment direction or reciprocally.

To get insights on the potential magnetic structures of the film, we have performed DFT simulations considering the two non-colinear magnetic structures that are proposed for V C$_{\rm{SO}}$ and V G$_{\rm{SO}}$ experimentally -- namely C$_x$C$_z$ and G$_y$ in Bertaut’s notation for a $Pbnm$ cell of with $a^-a^-c^+$ octahedral rotations, respectively. We then include the $4f$ Pr spins forming a C$_x$C$_z$ spin order as proposed in bulk compounds whatever the V spin orders or forced it to be ferromagnetically aligned along $z$ to mimic the Pr spin reorientation with an external magnetic field observed experimentally. The ground state corresponds to a C-type AFM ordering of the Pr$^{3+}$ magnetic moments associated to G-type V$^{3+}$ spin order~\cite{SM}. The average magnetic moment carried by height sites is close to $0.1\mu_{B}$ per 8 f.u. along the $[001]$ and $[110]$ directions, and around $-0.04\mu_{B}$ (8 f.u.)$^{-1}$ along the $[1\bar{1}0]$ direction for both V$^{3+}$ and Pr$^{3+}$ (see Table S3 in ref.~\cite{SM}). A magnetic configuration with a Pr$^{3+}$ C-type ordering is also found with C$_{\rm{SO}}$ ($\Delta{\rm E}=+10$~meV), whose net magnetization is rather small (see Table S4 in ref.~\cite{SM}). In Figure~\ref{Fig4}(d), we have sketched the average magnetic moments at just one of height sites in the supercell (blue vector for V$^{3+}$ at the center of a pseudo-cubic cell and orange vector for Pr$^{3+}$ at the corner of a pseudo-cubic cell) for the calculated magnetic configurations. Moreover, we have also calculated the case where the Pr$^{3+}$ spins are freezed along the $[110]$ direction (the orbital moment adapts to SOC), mimicking a ferromagnetic alignment (F-type) with \textbf{B}, before the supercells relaxation with either G$_{\rm{SO}}$ or C$_{\rm{SO}}$ V$^{3+}$ spin orders. Indirectly, we apply a magnetic field. Remarkably, only the C$_{\rm{SO}}$/F-type magnetic configuration yields a significant net magnetic moment along the $[110]$ direction for V$^{3+}$ with with $3.955~\mu_B$ (8 f.u.)$^{-1}$ and illustrated in Figure~\ref{Fig4}(d) (see Table S6 in ref.~\cite{SM}). Although the G$_{\rm{SO}}$ is more stable after ZFC, only the forced ferromagnetic alignment of the Pr$^{3+}$ with \textbf{B} and C$_{\rm{SO}}$ ordering of the V$^{3+}$ spins yield a net magnetization of along the $[110]$ direction, further rationalizing the V$^{3+}$ spins reordering. Experimentally, we have only observed an open hysteretic cycle for V$^{3+}$ when \textbf{B} was along the [110] direction. For Pr$^{3+}$, angle dependent measurements showed that $\langle\rm L_{\rm{z}}\rangle$ is constant at saturation, whereas it vanishes at remanence when \textbf{B} is along the $[100]$ or $[010]$ directions (see Figure S4 in ref.~\cite{SM}). This anisotropic behavior at remanence suggests that the magnetic polarizability of Pr$^{3+}$ depends also on crystalline structure. For instance, the off-centering of the Pr$^{3+}$ ion from its position in cubic symmetry and/or V$^{3+}$ $3d$ orbital ordering (entangled with spin orders) could lead to a change in the orbital sector, tuning $\langle\rm L_{\rm{z}}\rangle$ through perturbation of the single-ion anisotropy.

\subsection{Strain engineering.} 
\begin{figure*}[tp]
    \begin{center}
        \includegraphics[trim={3.5cm 1.0cm 2cm 0.2cm}, clip, width=0.87\textwidth]{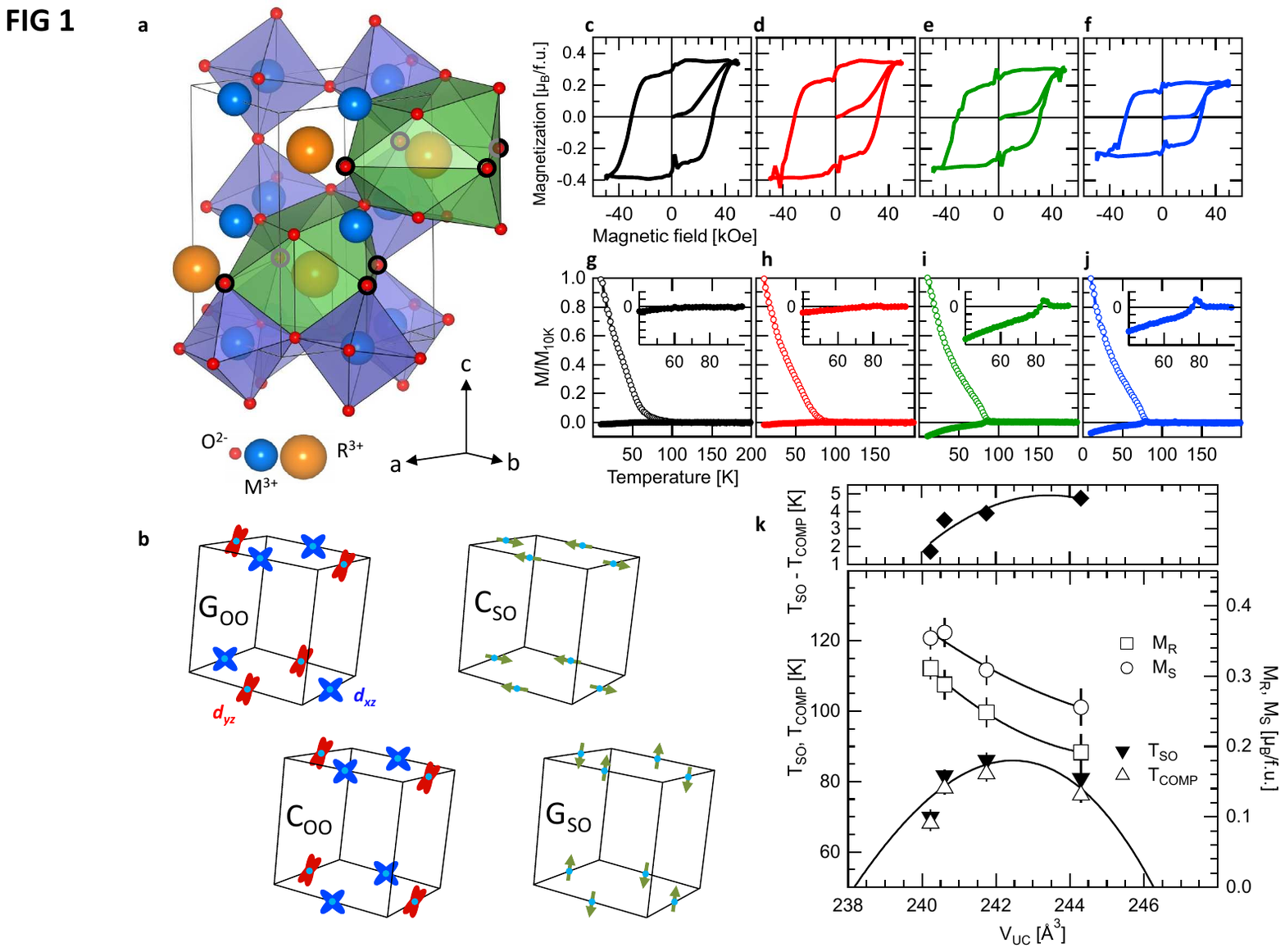}
    \end{center}
\caption{\textbf{Structure, orbital/spin orders of RVO$_3$, and PrVO$_3$ magnetic properties.} (a) Schematic of the orthorhombic structure of RVO$_3$. V$^{3+}$ and R$^{3+}$ ions are surrounded by 6 oxygen ions (octahedral cage in blue) and the R$^{3+}$ by 12 oxygen ions (cuboctahedral cage in green), respectively. Some octahedra and cuboctahedra are removed for clarity. (b) Schematics of the orderings of the V$^{3+}$ $3d_{xz}$ (blue) and $3d_{yz}$ (red) orbitals and the corresponding spin orderings (the $3d_{xy}$ orbital is not shown for clarity). Ordering with alternating occupied orbitals in all directions is labelled as G$_{\rm{OO}}$, while the one with identical occupation between consecutive planes along the $c$-axis, is labelled as C$_{\rm{OO}}$. The antiferromagnetic C$_{\rm{SO}}$  and G$_{\rm{SO}}$ spin orders are characterized by spins direction lying in the $ab$ plane and along the $c$-axis, respectively. (c)-(f) At 10~K, magnetization as a function of magnetic field applied parallel to the film's surface, for the films $\#1$, $\#2$, $\#3$ and $\#4$ grown at 800$^{\circ}$C (c), 700$^{\circ}$C (d), 600$^{\circ}$C (e) and 500$^{\circ}$C (f), respectively. (g)-(j) Respective normalized magnetization versus temperature after ZFC (full circles) and FC (empty circles). Insets: close-up of the ZFC curves around T$_{\rm{SO}}$. (k) Bottom panel, T$_{\rm{SO}}$ and T$_{\rm{COMP}}$ temperatures (left axis), and M$_{\rm{R}}$ and M$_{\rm{S}}$ magnetizations (rigth axis) as function of V$_{\rm{UC}}$. Top panel, difference of T$_{\rm{SO}}$ and T$_{\rm{COMP}}$. Lines are guides for the eyes.}
\label{Fig1} 
\end{figure*}

\begin{figure}[tp]
    \begin{center}
        \includegraphics[trim={3cm 5.4cm 7.5cm 3.3cm}, clip, width=0.45\textwidth]{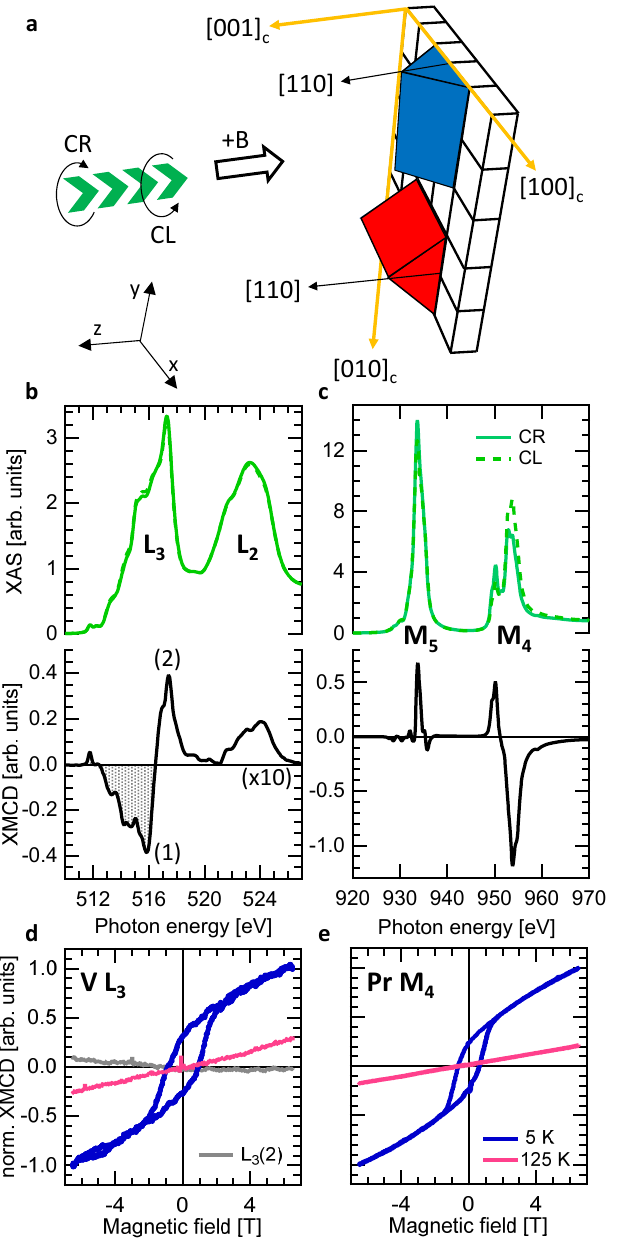}
    \end{center}
\caption{\textbf{XAS and XMCD at low temperatures.} (a) Schematic of the experimental geometry. The incident circularly polarized light, the magnetic field \textbf{B}, and the $[110]$ directions are parallel. (b) V-L$_{\rm{2,3}}$ edges and (c) Pr-M$_{\rm{4,5}}$ edges XAS and XMCD spectra, recorderd at T = 5~K with $|\textbf{B}|$ = 65~kOe. (1) and (2) indicate the position of V-L$^{(1)}_{3}$ (516~eV) and V-L$^{(2)}_{3}$ (517.5~eV), respectively. (d) and (e) XMCD hysteretic loops measured, at 5~K and 125~K, at V-L$^{(1)(2)}_{\rm{3}}$ edges and XMCD Pr-M$_{\rm{4}}$, respectively.}
\label{Fig2} 
\end{figure}

\begin{figure}[tp]
    \begin{center}
        \includegraphics[trim={5.5cm 8cm 5.5cm 4.5cm}, clip, width=0.5\textwidth]{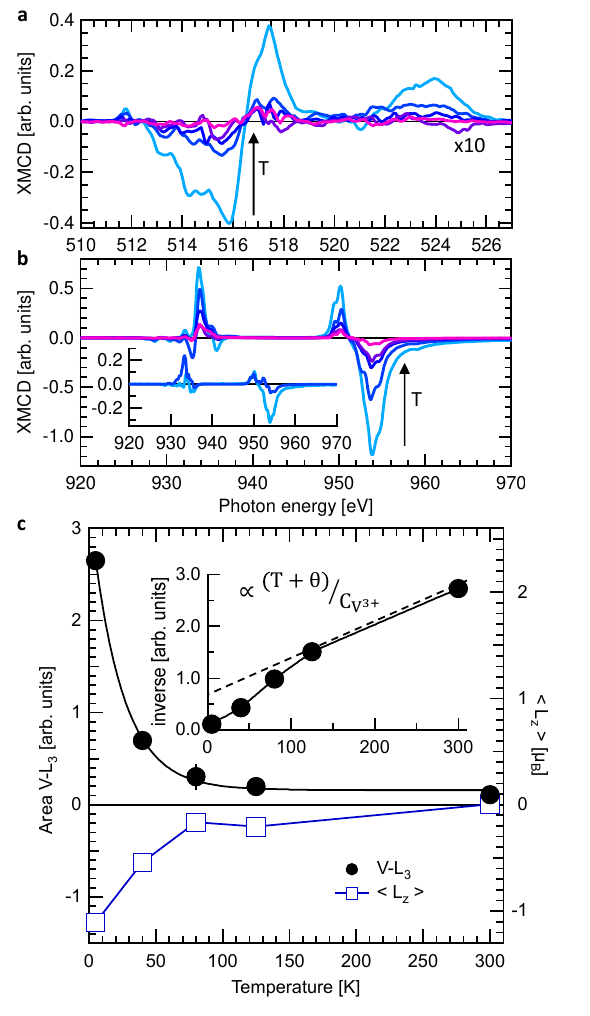}
    \end{center}
\caption{\textbf{XMCD spectra as a function of the temperature.} (a) V-L$_{\rm{2,3}}$ edges and (b) Pr-M$_{\rm{4,5}}$ edges XMCD spectra recorded at 5~K, 40~K, 80~K, 125~K and 300~K (light blue to violet lines) with $|\textbf{B}|$ = 65~kOe. The inset of (b) displays the remanent dichroic signal of recorded at 5~K and 40~K. (c) V-L$_{\rm{3}}$ area and Pr-M$_{\rm{4}}$ intensity versus temperature (lines are guide for the eyes). The inset displays the inverse of V-L$_{\rm{3}}$. The black dashed line is guide for the eyes.}
\label{Fig3} 
\end{figure}

\begin{figure*}[tp]
    \begin{center}
        \includegraphics[trim={4.5cm 4cm 4cm 1cm}, clip, width=0.9\textwidth]{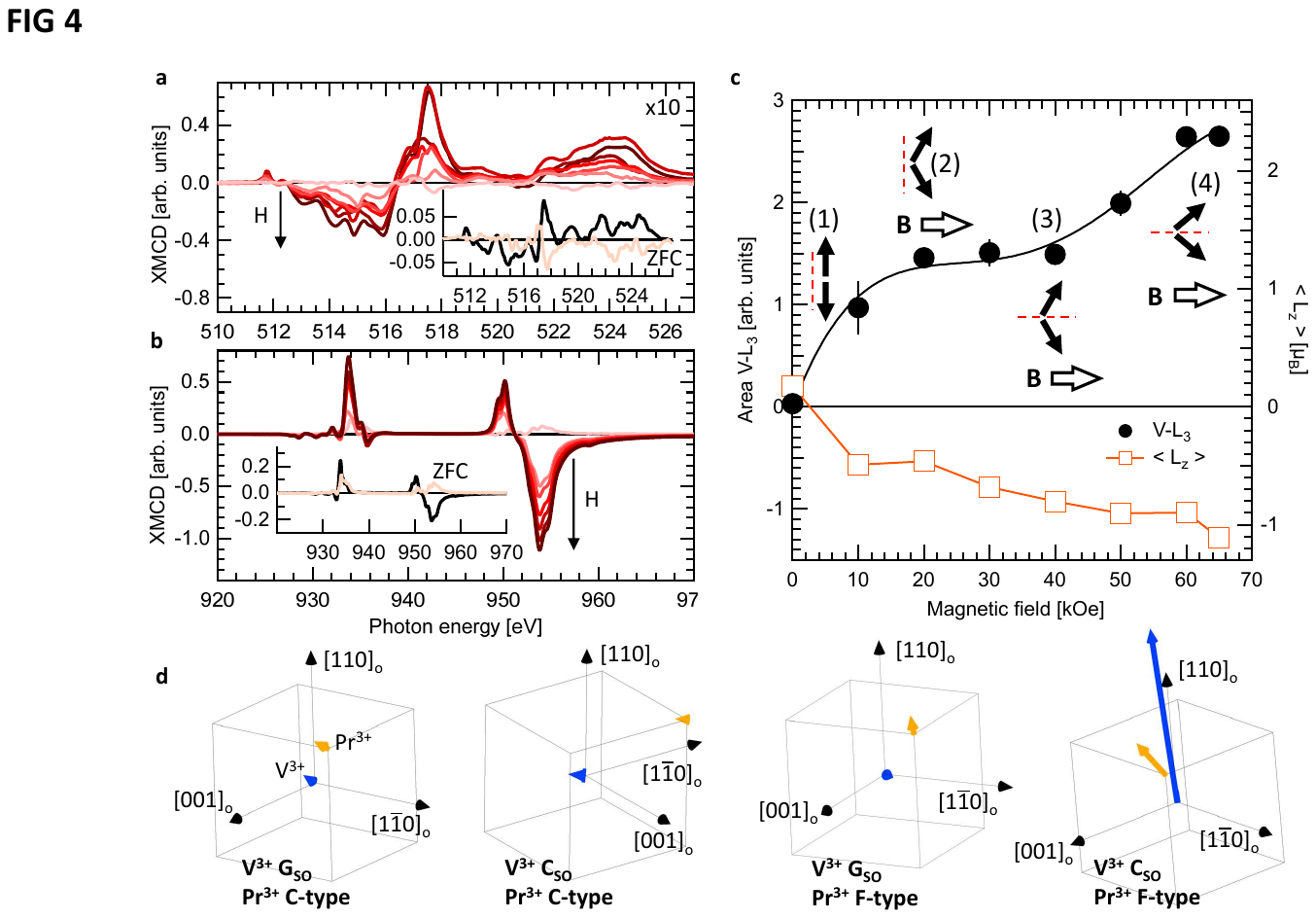}
    \end{center}
\caption{\textbf{XMCD as a function of B, at 5~K after ZFC.} (a) V-L$_{\rm{2,3}}$ edges and (b) Pr-M$_{\rm{4,5}}$ edges XMCD spectra. $|\textbf{B}|$ is varied from 0~Oe to 60~kOe (10~kOe step) up to 65~kOe (light brown to dark brown lines). The insets display the comparison between remanent (black) and ZFC (light brown) dichroic signal at V-L$_{\rm{2,3}}$ edges (a) and Pr-M$_{\rm{4,5}}$ edges (b). (c) V-L$_{\rm{3}}$ area and $\langle\rm L_{\rm{z}}\rangle$ versus \textbf{B} (lines are guide for the eyes). (d) sketchs of the calculated magnetic configurations of V$^{3+}$ and Pr$^{3+}$. In the right magnetic configuration, the blue vector's magnitude is divided by 3.}
\label{Fig4} 
\end{figure*}

\begin{figure*}[tp]
    \begin{center}
        \includegraphics[trim={3cm 11cm 3cm 2.5cm}, clip, width=1\textwidth]{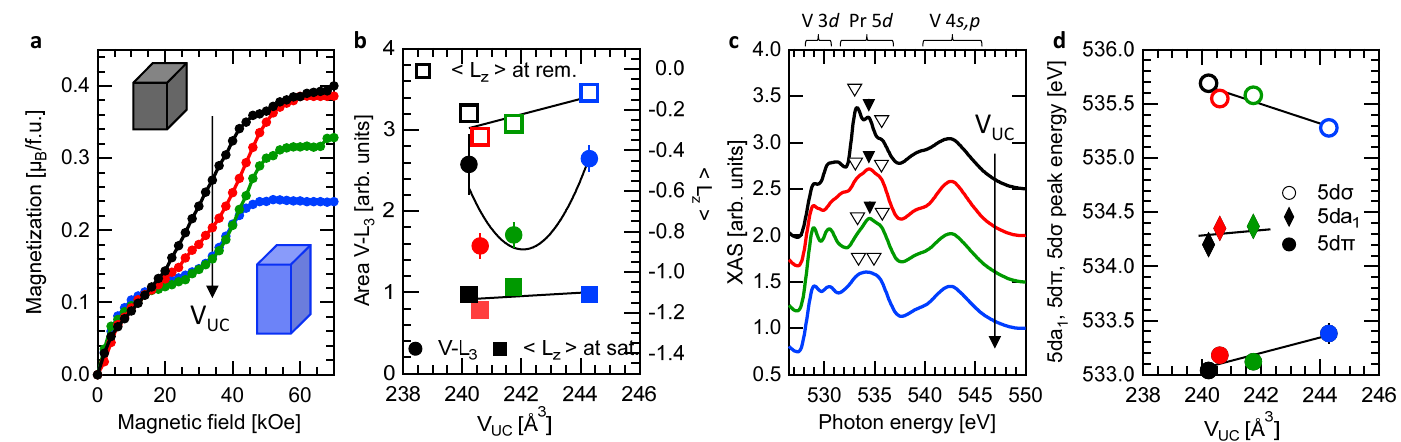}
    \end{center}
\caption{\textbf{Magnetic properties versus V$_{\rm{UC}}$, at 5~K.} (a) Macroscopic first magnetization curves measured for films ($\#1$ to $\#4$) with different V$_{\rm{UC}}$, with \textbf{B} perpendicular to the film's surface. (b) and V-L$_{\rm{3}}$ area (at saturation) and $\langle\rm L_{\rm{z}}\rangle$ (square), at remanence (open symbol) and at saturation (closed symbol), versus V$_{\rm{UC}}$. (c) O-K edge XAS spectra for films with different V$_{\rm{UC}}$. The peak energy of the V ($3d$ and $4sp$) and Pr ($5d$) orbitals hybridized with the O $2p$ orbitals are indicated. (d) Peak energy of the Pr $5d$ hybridized orbitals ($5d\sigma$, $5da_1$, $5d\pi$) as a function of V$_{\rm{UC}}$. The line is guide for the eyes. }
\label{Fig5} 
\end{figure*}

To gain more insight into the coupling mechanisms, we present in Figure~\ref{Fig5}(a) the first macroscopic magnetization curves as a function of V$_{\rm{UC}}$, measured with \textbf{B} applied along the $[110]$ direction. For small V$_{\rm{UC}}$ (black curve, $\#1$), the magnetization smoothly rises with \textbf{B} until the saturation. A kink around 20~kOe corresponds to the metamagnetic transition. Its curvature increases with V$_{\rm{UC}}$ so that the curve mirrors (blue curve, $\#4$) that of $\cal{A}_{\rm{V^{3+}}}$ in Figure~\ref{Fig4}(c), interpreted as the V$^{3+}$ spin reorientation from G$_{\rm{SO}}$ to C$_{\rm{SO}}$. 
In Figure~\ref{Fig5}(b), $\cal{A}_{\rm{V^{3+}}}$ (at saturation) and $\langle\rm L_{\rm{z}}\rangle$ (at remanence and saturation) are displayed as a function of V$_{\rm{UC}}$. Overall the amplitude of $\langle\rm L_{\rm{z}}\rangle$ decreases as V$_{\rm{UC}}$ increases, albeit slightly at saturation, and the non linear dependence and $\cal{A}_{\rm{V^{3+}}}$ versus V$_{\rm{UC}}$ recalls the dependence of T$_{\rm{SO}}$ in Figure~\ref{Fig1}(k). Thus, the behaviors of $\langle\rm L_{\rm{z}}\rangle$ and $\cal{A}_{\rm{V^{3+}}}$ at the microscopic scale agrees the macroscopic measurements in Figure~\ref{Fig1}. Additionally, normalized XMCD loops for films with small and large V$_{\rm{UC}}$ (see Figure S6 in ref.~\cite{SM}) are in agreement with hysteretic loops in Figure~\ref{Fig1}, especially for Pr-M$_{\rm{4,5}}$ edges. This raises the question of the change of the Pr$^{3+}$ cuboctahedral environment, as shown in Figure~\ref{Fig1}(a), its impact on magnetic properties. 
In Figure~\ref{Fig5}(c), we show the XAS spectra recorded at the O-K, varying V$_{\rm{UC}}$ of the films. The observed features correspond to the transition from O $1s$ to O $2p$ orbitals, hybridized with the empty orbital states of the V and Pr cations, thereby sensitive to their surrounding environment. The intensity of the peaks increases as V$_{\rm{UC}}$ decreases, consistently with a higher hybridization in small unit cell. The broad peak around 542.5~eV and the peaks around 535~eV correspond to the V 4$sp$ and Pr $5d$ hybridized orbitals, respectively. The peaks at 529~eV and 531~eV, partially obscured by additional peaks for small V$_{\rm{UC}}$, correspond to the O $2p$ overlap with the crystal field split V $3d$ $t_{2g}$ and $e_g$ orbitals, respectively.
In Figure~\ref{Fig5}(d), we show, against V$_{\rm{UC}}$, the energy of the O~$2p$ $-$ Pr~$5d$ hybridized peaks, whose bonding characters, in a quasitrigonal environment, correspond to $\sigma$-bonds ($5d\sigma$), $\pi$-bonds ($5d\pi$) and Pr~$5d_{z^2}$ bonds ($5da_1$) with O $2p$ character mixed in~\cite{Cao2016}. The Pr~$5da_1$ shows a slight shift, and its intensity vanishes as V$_{\rm{UC}}$ increases. The Pr~$5d\pi$ and Pr~$5d\sigma$ peaks shift towards higher and lower energy, respectively. We attribute these shifts to the elongation of the O~$2p$ $-$ Pr~$5d\sigma$ bond and shrinking of the O~$2p$ $-$ Pr~$5d\pi$ bond with increasing V$_{\rm{UC}}$. 

We used DFT calculations to assess the structural changes of PrVO$_3$ under strain along the $[110]$ direction and the evolution of the Pr$-$O bond lengths (see above). The bonds increases (up to 6\%) for Pr atoms bound with all apical O atoms of the VO$_6$ octahedron, while they decreases (approx. $-1.5\%$) for Pr atoms bound with some equatorial O atoms, especially along the $[110]$ direction (see Figure S7 in ref.~\cite{SM}). Thus, this suggests that the highly directional $5d\sigma$ bonds are sensitive to an elongation involving apical oxygen [highlighted in Figure~\ref{Fig1}(a)] and that the shrinking of the $5d\pi$ bonds involves oxygen in the equatorial plane of the VO$_6$ octahedra. The measured energy shifts hence indirectly capture the distortion of the VO$_6$ octahedra through changes of the cuboctahedral environment of Pr$^{3+}$. 
Recently, theoretical investigations have highlighted the role played by the Dzyaloshinski-Moriya interactions and magnetostructural couplings involving the oxygen octahedra rotation in the understanding of the polarization of the R$^{3+}$ magnetic moment in orthorhombic perovskites~\cite{Zhao2016b,Sasani2021}. For instance, Zhao \textit{et al.} have considered an energy term as $\Delta E_{\rm{R^{3+}}}\propto-(\mathbf{\omega_{-}}\times \mathbf{G_{\rm{SO}}}+\mathbf{\omega_{+}}\times \mathbf{C_{\rm{SO}}}).\mathbf{F^{\rm{R^{3+}}}}$ among the magnetic interactions between the cations~\cite{Zhao2016b}. $\mathbf{F^{\rm{R^{3+}}}}$ is the net magnetization of R$^{3+}$, $\mathbf{\omega}$ is a pseudo-vector representing in-phase ($+$) or out-of-phase ($-$) octahedral rotations, and $\mathbf{C_{\rm{SO}}}$/$\mathbf{G_{\rm{SO}}}$ is a vector representing the spin order of the transition metal~\cite{Bellaiche2012}. In the parent compound LaVO$_3$, which is also (110)-oriented when grown on SrTiO$_3$, the $a^-a^+c^-$ tilt system has been determined experimentally~\cite{Rotella2012}. This indicates that the in-phase rotation axis lies in the substrate plane, as deduced indirectly in PrVO$_3$ and LaVO$_3$ films by mapping the R$^{3+}$ cation displacements~\cite{Masset2020}, which are coupled to the rotations through a trilinear energetic term~\cite{Amisi2012,Mio2013}. Thereby, the V$^{3+}$ spin is G$_{\rm{SO}}$ in our case after ZFC, and $\mathbf{\omega_{-}}$ and $\mathbf{\omega_{+}}$ are parallel to the $[010]$ and $[001]$ directions, respectively. From Figure~\ref{Fig1}(a), we consider the V$^{3+}$ spins align along $c$ for G$_{\rm{SO}}$ and within the $ab$ plane for C$_{\rm{SO}}$ (assuming the $b$-component is the strongest as in the bulk material~\cite{Reehuis2016}). Consequently, we can write $\Delta E_{\rm{R^{3+}}}\propto-(|\mathbf{\omega_{-}}|.|\mathbf{G_{\rm{SO}}}|+|\mathbf{\omega_{+}}|.|\mathbf{C_{\rm{SO}}}|).|\mathbf{F^{\rm{R^{3+}}}}|$ along the $[100]_o$ direction, for simplicity. We observe that depending on the $\mathbf{F^{\rm{R^{3+}}}}$ direction, the first or the second term of $\Delta E_{\rm{R^{3+}}}$ will gain energy, with $\mathbf{F^{\rm{R^{3+}}}}$ along $[100]$ or $[\bar{1}00]$, respectively. In turn, it will promote a spin V$^{3+}$ reorientation. As \textbf{B} contributes to align the Pr$^{3+}$ magnetic moments, it will induce a spin reodering from G$_{\rm{SO}}$ to C$_{\rm{SO}}$ as in DyVO$_3$~\cite{Fujioka2010}, and promote a net magnetic moment on the V$^{3+}$ sublattice as observed in Figure~\ref{Fig4}. Hence, tailoring the octahedral site rotations and distortions should allow tuning the $3d-4f$ exchange interaction and the energy balance between the different ordered states.

\subsection{Tailoring distortions and exchange couplings.} 
\begin{figure*}[tp]
   \begin{center}
        \includegraphics[trim={2cm 3.5cm 2.5cm 2.5cm}, clip, width=1\textwidth]{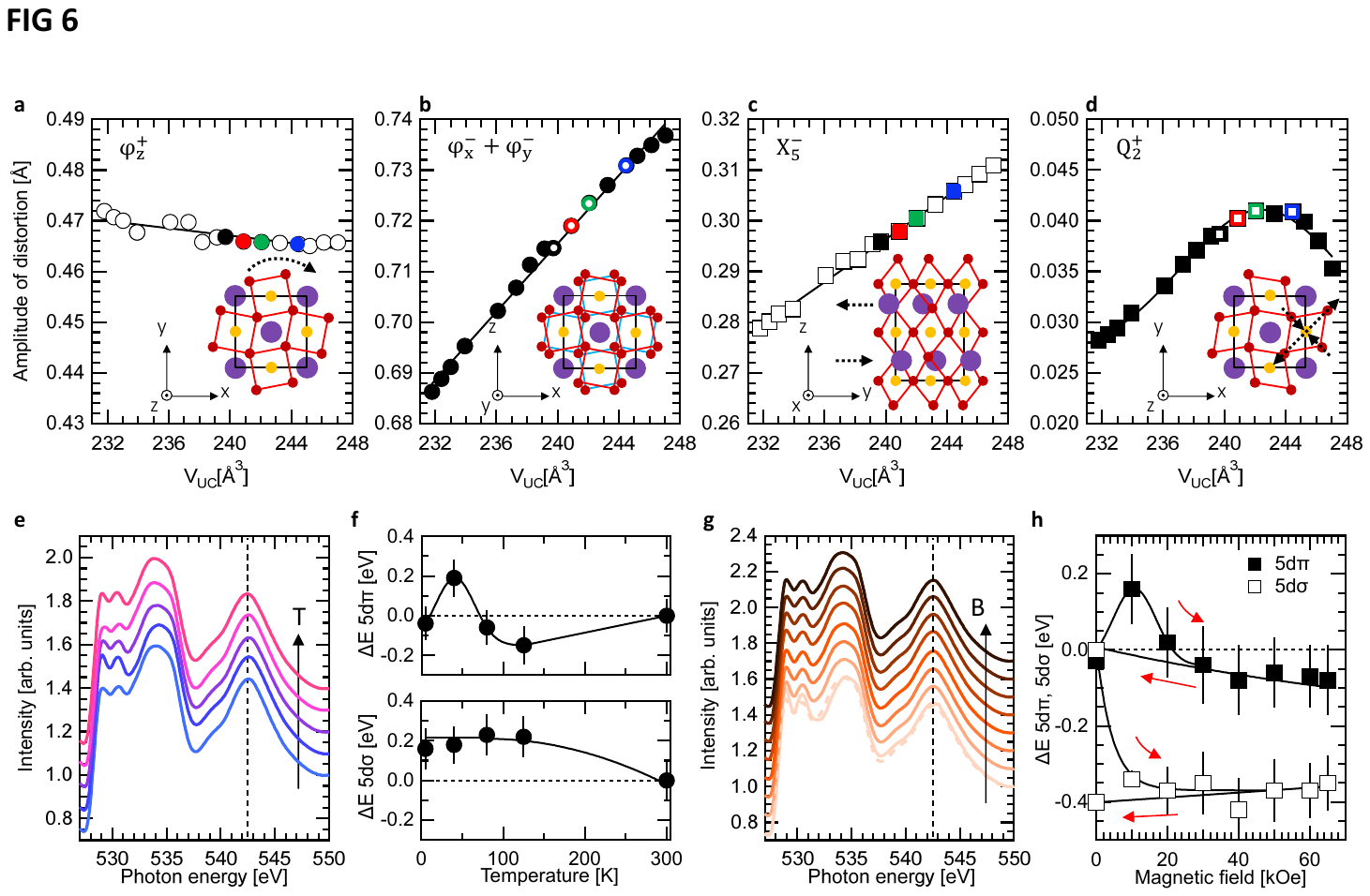}
   \end{center}
\caption{\textbf{Lattice distortions and O 2p $-$ Pr 5d orbitals hybridization.} (a)-(d) $\varphi_{z}^{+}$, $\varphi_{xy}^-$ ($\varphi_x^-+\varphi_y^-$), $X_5^-$ and $Q_2^+$ calculated lattice distortions versus V$_{\rm{UC}}$, respectively. The calculated data corresponding to the deposited films ($\#1$ to $\#4$) are marked with colored symbol. (e) O-K edge XAS spectra versus temperature, with $|\textbf{B}|=65$~kOe. (f) Peak energy shift (with respect to the spectrum at 300~K) of the Pr $5d\sigma$ and $5d\pi$ hybridized orbitals versus temperature, with $|\textbf{B}|=65$~kOe. (g) At 5~K and after ZFC, O-K edge XAS spectra versus \textbf{B}. $|\textbf{B}|$ is varied from 0~Oe to 60~kOe (10~kOe step) up to 65~kOe (light brown to darck brown lines). The measurement at remanence is show in dashed light brown line. (h) Peak energy shift (with respect to the spectrum after ZFC) of the Pr $5d\sigma$ and $5d\pi$ hybridized orbitals versus \textbf{B}. The line is guide for the eyes.}
\label{Fig6} 
\end{figure*}

To gain further insights, we calculated the amplitude of the distortion modes in PrVO$_3$ films as a function of the elongation along the $[110]$ direction by DFT (see above and in ref.~\cite{SM}). Four primary distortions are active: two octahedral rotations $\varphi_{z}^{+}$, $\varphi_{xy}^-$ ($\varphi_x^-+\varphi_y^-$), an antipolar motion of R$^{3+}$ cations $X_5^-$ and a JT mode $Q_2^+$~\cite{Varignon2015}. Their amplitude as function of V$_{\rm{UC}}$ is presented in Figure~\ref{Fig6}(a)--(d), respectively. The $[001]$ axis of in-phase rotation lies in the substrate plane, resulting in a minimal $\varphi_{z}^{+}$ rotation due to substrate clamping [Figure~\ref{Fig6}(a)]. In contrast, the $[110]$ axis of the out-of-phase rotation is free, and the $\varphi_{xy}^-$ rotation increases with V$_{\rm{UC}}$ [Figure~\ref{Fig6}(b)]. This increase is known to favor a C$_{\rm{OO}}$ ground state in bulk RVO$_3$~\cite{Sage2007} but it promotes also the anti-polar $X_5^-$ motion~\cite{Goodenough1971,Mizokawa1999} of the Pr$^{3+}$ ions in the $ab$ plane [Figure~\ref{Fig6}(c)]. We note the $[110]$ axis elongation appears then as an appealing means for enhancing the electric polarization driven by hybrid improper ferroelectricity~\cite{Rondinelli2012,Varignon2015,Spaldin2007,Bousquet2008,Benedek2011}. The $X_5^-$ motion forces the appearence of the $Q_2^+$ distortion mode [Figure~\ref{Fig6}(d)], but exhibiting a non-monotonous behavior versus V$_{\rm{UC}}$ that is reminiscent of the trends in T$_{\rm{SO}}$ and $\cal{A}_{\rm{V^{3+}}}$ displayed in Figure~\ref{Fig1}(k) and Figure~\ref{Fig5}(c), respectively. This indicates that the superexchange interactions between V sites are mainly driven by the $Q_2^+$ octahedral-site distortion (linked to the octahedral rotation and cation displacements~\cite{Varignon2015}) in our films. Consequently, it promotes a C$_{\rm{OO}}$ ground state, associated with G$_{\rm{SO}}$. The strengthening of $\varphi_{xy}^-$ and $Q_2^+$ as V$_{\rm{UC}}$ increases should stabilize further a C$_{\rm{OO}}$/G$_{\rm{SO}}$ ground state. However, $Q_2^+$ is destabilized for large V$_{\rm{UC}}$, suggesting alternative means for energy gain. The decrease of the $Q_2^+$ distortion may suggest a potential larger energy gain associated to the $Q_2^-$ Jahn-Teller distortion and its G$_{\rm{OO}}$ ordering as shown in bulk perovskite oxides~\cite{Mizokawa1999,Zhou2006,Varignon2019}. Thus, the response of the system when applying \textbf{B}, forcing the V$^{3+}$ spins alignment in the $ab$ plane, will be the promotion of C$_{\rm{SO}}$, and concomitantly a G$_{\rm{OO}}$/$Q_2^-$ state to gain energy. We point out this should also favor an alternative antipolar $X_3^-$ motion of the Pr$^{3+}$ cations along the $c$-axis~\cite{Varignon2015}. The lattice distortions act thus as a tuning knob enabling the competition between C$_{\rm{SO}}$/G$_{\rm{OO}}$ and G$_{\rm{SO}}$/C$_{\rm{OO}}$ states, as seen in Figure~\ref{Fig5} where G$_{\rm{SO}}$ resists until the spins switch to C$_{\rm{SO}}$, as V$_{\rm{UC}}$ increases. In turn, octahedral-site JT modifications will also yield changes in the cuboctahedral environment surrounding Pr$^{3+}$. 

We have seen previously that structural changes due to the lattice expansion affect the hybridization between orbitals [see Fig.~\ref{Fig5}(c) and (d)]. We know probe these changes when the temperature or \textbf{B} are varied. Figure~\ref{Fig6}(e) and (g) display the O-K edge XAS spectra as a function of the temperature ($|\textbf{B}|=65$~kOe) and  the magnetic field $\textbf{B}$ at 5~K after ZFC, respectively. Figure~\ref{Fig6}(f) and (h) show the corresponding energy shifts of the O~$2p$ $-$ Pr~$5d\sigma$ and O~$2p$ $-$ Pr~$5d\pi$ hybridized peaks relative to 300~K and zero field, respectively. Notably, the energy of the O~$2p$ $-$ V 4$sp$ peak, indicated by a dashed line, remains constant at 542.55~$\pm$0.03~eV, during the temperature or magnetic field variation. The $5d\sigma$ peak shift towards higher energies in Figure~\ref{Fig6}(f) bottom panel and indicates that the length of $\sigma$-bonds decreases due to the volume contraction as the temperature decreases. The $5d\pi$ peak exhibits energy shifts [see Figure~\ref{Fig6}(f) top panel], indicating an initial $\pi$-bonds elongation, a subsequent shrink below 100~K, and a further elongation below 40~K. The shift at 100~K indicates the Pr$^{3+}$ magnetic sublattice polarization with the onset of the V$^{3+}$ spin ordering around T$_{\rm{SO}}$, likely reminiscent of magnetostriction. As a reversal of $\langle\rm L_{\rm{z}}\rangle$ at remanence occurs between 5 and 40~K [Figure~\ref{Fig3}(b)], these shift around 40~K indicates the modifications of octahedral-site distortions, likely reminiscent orbital order switching. 
Figure~\ref{Fig6}(h) shows the $5d\pi$, at 5~K (after ZFC), initially shifting towards high energy up to 10~kOe, then decreasing and returning back to its initial energy position at remanence. In contrast, the $5d\sigma$ peak energy drops by $-0.4$~eV, maintaining this energy position even at remanence. We attribute the change of the O $2p$ $-$ Pr $5d$ hybridization to the modifications of the VO$_6$ octahedral environment (see ref.~\cite{SM} for DFT calculations at the collinear level) with the reorientation of the V$^{3+}$ spins from G$_{\rm{SO}}$ to C$_{\rm{SO}}$ (rotation of the N\'eel vector), accompanied by a reversal of $\langle\rm L_{\rm{z}}\rangle$ observed in Figure~\ref{Fig4}. However, the behavior of the shift of the Pr $5d$ peak's energy is not entirely reversible with \textbf{B} and it suggests a pinning to atomic disorder or the mutual polarization between the Pr$^{3+}$ and V$^{3+}$ magnetic sublattices. Consequently, an exchange strictive structural bias may foster the coexistence of multiple states~\cite{Sage2007,Fujioka2010} and/or contribute to the observed large coercive fields as structural distortion, V$^{3+}$ spin reorientation and Pr$^{3+}$ magnetic moments are interlocked. Finally in the case of coexisting orbital orders, $Q_2^+$ and $Q_2^-$ JT distortions would be active simultaneously, both coupled to $X_3^-$~\cite{Varignon2015}. The resulting trilinear coupling has been identified as a key structural mode for controlling spins with an electric polarization in designed superlattices, paving the way towards orthorhombic perovskite based magnetoelectric multiferroics~\cite{Varignon2015}.   

\section{Conclusion}
In conclusion, we have shown that magnetization reversal and spin reorientation in PrVO$_3$ thin films involve the coupling between the antiferromagnetic V$^{3+}$ and paramagnetic Pr$^{3+}$ sublattices, with the canted $3d$ spins and $4f$ moments compensating each other as a function of the temperature. Through uniaxial strain engineering, we evidenced that the lattice distortions act on the exchange interaction between V$^{3+}$ and Pr$^{3+}$ cations, and consequently have a strong influence on the V$^{3+}$ spin and orbital orderings, assisted by a change of the cooperative Jahn-Teller distortion mode. Associated to the rare earth atomic displacements $X_5^-$ and $X_3^-$ in perovskite oxides, the consideration of these key magnetostructural coupling between A and B cations opens up new insights for the design of robust magnetoelectric multiferroics. The nature of the A-site atom, its single-ion anisotropy  and spin-orbit coupling should also be considered systematically for the understanding of the complexity in TMO.
\begin{acknowledgements}
OC thanks B. Raveau and M. Seikh for stimulating discussions. This work was supported by the French National Research Agency (ANR) under project CITRON (ANR-21-CE09-0032), the Région Grand Est under project RHUM (AAP-013-075), and the France 2030 government investment plan under grant PEPR SPIN - SPINMAT (ANR-22-EXSP-0007). The work has benefeted of the resources of experimental platforms at IJL: Tube Davm (funded by FEDER EU, ANR, Région Grand Est, and Métropole Grand Nancy), Magnetism \& Cryogenics, and XGamma, all supported by the LUE-N4S project, part of the French PIA project Lorraine Université d’Excellence (ANR-15IDEX-04-LUE), and by FEDER-FSE Lorraine et Massif Vosges 2014-2020, an EU program. The XAS experiments were performed on the DEIMOS beamline at SOLEIL Synchrotron (proposal number 20130636). We are grateful to the SOLEIL staff for smoothly running the facility. The first-principles calculations were conducted on the Abel supercomputers (PRACE project TheoMoMuLaM), the Cartesius supercomputer (PRACE project TheDeNoMO), and the Tier-1 supercomputer of the Fédération Wallonie-Bruxelles, funded by the Walloon Region (Grant No 1117545). Additional resources were provided by the Céci-HPC facilities, funded by F.R.S.-FNRS (Grant No 2.5020.1). JV acknowledges access granted to High Performance Computing resources of CRIANN through the projects 2020005 and 2007013 and of CINES through the DARI project A0080911453.  
\end{acknowledgements}
%
%
%\bibliography{biblio}
%

%

\end{document}